\documentclass[aps,preprint,nofootinbib,floatfix]{revtex4-1}
\usepackage{graphicx,color}
\usepackage[latin1]{inputenc}
\usepackage{amsmath,amssymb}
\usepackage{hyperref}
\usepackage{epstopdf}
\newcommand{\lsim}{\mathrel{\mathop{\kern 0pt \rlap
  {\raise.2ex\hbox{$<$}}}
  \lower.9ex\hbox{\kern-.190em $\sim$}}}
\newcommand{\gsim}{\mathrel{\mathop{\kern 0pt \rlap
  {\raise.2ex\hbox{$>$}}}
  \lower.9ex\hbox{\kern-.190em $\sim$}}}

\begin{document}
\title{Global Study of the Simplest Scalar Phantom Dark Matter Model}
\author{
Kingman Cheung$^{1,2}$, Yue-Lin S. Tsai$^3$, Po-Yan Tseng$^{2}$,\\
Tzu-Chiang Yuan$^4$ and A. Zee$^{4,5}$
}

\affiliation{
$^1$Division of Quantum Phases \& Devices, School of Physics, 
Konkuk University, Seoul 143-701, Korea\\
$^2$Department of Physics, National Tsing Hua University, 
Hsinchu 300, Taiwan\\
$^3$National Center for Nuclear Research, Hoza 69, 00-681 Warsaw, Poland\\
$^4$Institute of Physics, Academia Sinica, Nankang, Taipei 11529, Taiwan\\
$^5$Kavli Institute for Theoretical Physics, University of California, Santa Barbara, CA 93106
}

\date{\today}

\begin{abstract}
  We present a global study of the simplest scalar phantom dark matter
  model.  The best fit parameters of the model are determined by
  simultaneously imposing (i) relic density constraint from WMAP, 
  (ii) 225 live days data from direct experiment XENON100, 
  (iii) upper limit of gamma-ray flux from Fermi-LAT indirect detection 
  based on dwarf spheroidal satellite galaxies, and 
  (iv) the Higgs boson candidate with a mass about 125 GeV and 
  its invisible branching ratio no larger than 40\% if the
  decay of the Higgs boson  into a pair of dark matter is kinematically
  allowed.  The allowed parameter space is then used to
  predict annihilation cross sections for gamma-ray lines,
  event rates for three processes mono-$b$ jet, single charged
  lepton and two charged leptons plus missing energies at the Large
  Hadron Collider, as well as to evaluate the muon anomalous magnetic
  dipole moment for the model.
\end{abstract}
\maketitle

\section{Introduction}

Evidences for the existence of dark matter are mainly coming from
cosmological observations related to the physics of gravity. These include
the relic density of dark matter, anisotropies in the Cosmic Microwave
Background (CMB), large scale structure of the universe, as well as
the bullet clusters and the associated gravitational lensing effects.
While we still do not know what the nature of dark matter is, it is
clear that there is no room to accommodate dark matter in the standard
model (SM) of particle physics based on gauge invariance of $SU(3)_C
\times SU(2)_L \times U(1)_Y$ and Einstein-Hilbert gravity theory
based on general coordinate invariance.  While it is plausible that
the nature of dark matter may have a purely gravitational origin,
theories that have been put forward thus far are not as convincing as
those from the particle physics point of view. In particular the relic
density strongly suggests that dark matter may be a weakly interacting
massive particle (WIMP). If dark matter can indeed be related to weak
scale physics, there may be hope for us to detect them in various
underground experiments of direct detection as well as in space
experiments using balloons, satellites, or space station of indirect
detection.  Furthermore, WIMP dark matter might be produced directly
at the Large Hadron Collider (LHC) by manifesting itself as missing
energy with a spectrum that may be discriminated from standard model
background of neutrinos.

In this paper, we will focus on the simplest dark matter model
\cite{SZ} which is based on adding a real singlet scalar field to the
SM. The communication between the scalar dark matter 
and the SM gauge bosons and fermions must then
go through the SM Higgs boson.  While there have been many studies for
this simple model and its variants in the literature \cite{mcdonald,
burgess,hegang,michael,nmsm,Drozd:2011aa}, we believe a global study
of this model is still missing.  In this work, we will fill this
gap. We use the current experimental constraints of relic density from
WMAP \cite{wmap}, 225 live days data from direct experiment XENON100
\cite{xenon2012}, diffuse gamma-ray flux from indirect detection
experiment of Fermi-LAT using the dwarf spheroidal satellite galaxies
(dSphs) \cite{fermilat-dSphs,GSK-dSphs}, and a Higgs boson candidate with 
mass about 125 GeV reported recently by the LHC \cite{atlas,cms} to deduce the best fit
parameters of the model.  The deduced parameters are used to predict
various phenomenology of the model at the LHC, including production of
the mono-$b$ jet, single charged lepton, and two charged leptons plus
missing energies.  We also evaluate the muon anomalous magnetic dipole moment
which is a two loop process in the model. 
For a global fitting based on effective operators
approach, see our recent work in \cite{globaleffop}. A similar global analysis 
for isospin violating dark matter is presented in \cite{globalivdm}.

In the next section, we will briefly review the scalar phantom model of dark matter. In section III, 
we present the global fitting for the relevant parameters of the model using 
the various experimental constraints described above.
In section IV, we discuss collider phenomenology and the muon 
anomalous magnetic dipole moment of the model. We conclude in section V.
Some analytical formulas of the matrix elements needed in our analysis 
as well as the expression for the muon anomalous magnetic dipole 
moment are collected in the Appendix.

\section{The Scalar Phantom Model}

The simplest dark matter model (SZ) \cite{SZ}
(dubbed scalar phantom by the authors in \cite{SZ}) 
is obtained by adding one real singlet scalar $\chi$ 
in addition to the Higgs doublet $\Phi$ to the SM.
The scalar part of the  Lagrangian is given by
\begin{equation}
\label{SZ}
{\cal L}_{\rm scalar} = \left( D^\mu \Phi \right)^\dagger \left( D_\mu \Phi \right)
-  \lambda \left( \Phi^\dagger \Phi - \frac{\mu^2}{2 \lambda} \right)^2
+ \frac{1}{2}\partial^\mu \chi \partial_\mu \chi   
- \frac{1}{2} m^2 \chi^2
-\frac{1}{4!} \eta \chi^4
-\frac{1}{2}\rho \chi^2 \Phi^\dagger \Phi \;.
\end{equation}
A discrete $Z_2$ symmetry of $\chi \to - \chi$ while keeping all SM 
fields unchanged
has been imposed to eliminate the $\chi$, $\chi \Phi^\dagger \Phi$, 
and $\chi^3$ terms. As a result it guarantees 
the stability of the $\chi$ particle and hence it may be
a viable candidate for WIMP (weakly interacting massive particle) dark matter.
Note that the $\chi^4$ term in Eq.(\ref{SZ}) implies a contact interaction vertex among the scalar dark matter.

The virtue of this model is its simplicity. Indeed, it represents the
simplest realization of a broad 
class of models, in which we could add
any number of singlet scalar $\chi$ to the standard model, or the
standard model augmented by a private Higgs sector \cite{PZ}. The
analysis given here is in the spirit of seeing whether or not the
simplest version of this kind of model could now be ruled out.

After electroweak symmetry breaking, $\Phi$ develops a vacuum expectation value
$v/\sqrt{2}$, where 
$v= \mu/\sqrt{\lambda} = 246$ GeV. 
After making the shift $\Phi (x)^T = \left( 0 \; ,\; v + H(x) \right) / \sqrt 2$, 
the physical Higgs field $H$ obtains a mass $m_H = \sqrt{2 \lambda} v = \sqrt 2 \mu$ and 
the last term 
in Eq.(\ref{SZ}) becomes
\begin{eqnarray}
\label{shift}
-\frac{1}{2}\rho \chi^2 \Phi^\dagger \Phi
& \longrightarrow & 
-\frac{1}{4}\rho v^2 \chi^2 
- \frac{1}{2}\rho v H \chi^2
- \frac{1}{4} \rho H^2 \chi^2 \; \; .
\end{eqnarray} 
The first term on the right handed side of Eq.(\ref{shift}) implies the 
dark matter $\chi$ also pick up an additional contribution 
of $ \frac{1}{2}\rho v^2$ to its mass, thus
$m_\chi^2 = m^2 + \frac{1}{2}\rho v^2$. We will assume $m_\chi^2$ is always positive
so that the $Z_2$ symmetry will never be broken, except perhaps due to black hole effects.
The second term in Eq.~(\ref{shift}) tells us that the dark matter $\chi$ 
can communicate to the SM fields and self-interact with itself 
via a tree level Higgs exchange, while 
the last term contributes to the relic density calculation 
from the process $\chi\chi \to HH$ if kinematically allowed.
If kinematics permits, the second term
also allows Higgs boson to decay into a pair of $\chi$, giving rise to the invisible Higgs width.
Implication of invisible Higgs width in the Higgs search at the LHC will be discussed further
in the following sections.

There are a few theoretical restrictions on the model, including
vacuum stability, unitarity, and triviality.
Stability of the vacuum requires the scalar potential be bounded from below. At tree level, we have
\begin{equation}
\label{paracons-1}
\lambda > 0 \; , \; \eta  > 0 \; , \; \rho^2 < \frac{2 \lambda \eta }{3} \; .
\end{equation}
Tree level perturbative unitarity constraints can be deduced by considering the 
longitudinal gauge boson scatterings \cite{vvvv} as well as all
scalar-scalar scatterings \cite{michael}
\begin{equation}
m_H^2 < \frac{8 \pi }{3} v^2  \approx \left( 712 \,\rm GeV \right)^2 \; , 
\; \;\; \eta < 8 \pi \;, \;\; {\rm and} 
\;\; \vert \rho \vert < 8 \pi \; .
\end{equation}
Analysis of the triviality of this model can be found in the 
literature \cite{michael,nmsm}.

Self-interacting cold dark matter was proposed in
\cite{Spergel-Steinhardt} to resolve some conflicts between actual
observations and WIMP theory which predicts overly dense cores in the
center of galaxies and clusters and an overly large number of halos
within the Local Group.  The Spergel-Steinhardt bound
\cite{Spergel-Steinhardt} for collisional (self-interacting) dark
matter,
\begin{equation}
2 \times 10^3 \; {\rm GeV}^{-3} \leq \frac{\sigma_{\chi \chi \to \chi \chi}}{m_\chi} \leq 3 \times 10^4 \; {\rm GeV}^{-3} \;,
\end{equation}
can be used to constrain the contact self-coupling $\eta$ of the scalar phantom 
(as well as the coupling $\rho$ through Higgs exchange).
We refer to previous works \cite{burgess,Holz:2001cb} on this issue.

\section{Global Fitting}

In this section, we consider the global constraints coming from WMAP
relic density \cite{wmap}, the XENON100 data \cite{xenon2012}, the
Fermi-LAT upper limit of diffuse gamma-ray flux based on dSphs
\cite{fermilat-dSphs}, the LHC Higgs mass around 125 GeV
\cite{atlas,cms}, and the upper limit of Higgs invisible decay
branching ratio \cite{Djouadi-etal}. The relevant parameters of the
model are the dark matter mass $m_\chi$, the coupling $\rho$ and the
Higgs mass $m_H$, which the likelihood functions will depend on.  The
publicly available software package MicrOMEGAs version 2.4.5
\cite{micromegas} is used to calculate the relic density and gamma-ray
flux.  

\subsection{WMAP Relic Density}

In the calculation of the relic density, we consider all of
the following two body tree level processes
\begin{equation}
\label{tree-process}
\chi \chi \to H^{(*)} \to f \bar f,  \; W^+W^-, \; ZZ, \; HH \;\; .
\end{equation}
The relic density from the WMAP 7-year result \cite{wmap} is
\begin{equation}
\Omega_c h^2 = 0.112 \pm 0.0056 \; ,
\end{equation}
where $\Omega_c$ is the density of the cold dark matter normalized to the critical density
and $h$ is the Hubble rate in unit of 100 km sec$^{-1}$ Mpc$^{-1}$.
In the fitting, we use the Gaussian distribution 
for the WMAP relic density in our likelihood function, 
\begin{equation}
\label{eq:rd}
\mathcal{L}_{\rm relic}=e^{-\frac{\chi^2}{2}} \; ,
\end{equation} 
with the $\chi^2$ defined as \footnote{We apologize for abusing our notation of $\chi$ 
to stand for the scalar phantom dark matter as well as to define the chi-squared.}
\begin{equation}
\chi^2=\frac{\left(\rm{prediction}-\rm{experimental\,central\,value}\right)^2}{\sigma^2+\tau^2},
\end{equation}
where the $\sigma$ can be read off from the WMAP experimental error \cite{wmap}.
We also assume a theoretical uncertainty $\tau$ to be 10\% of the  
prediction in order to account for the discrepancy due to 
different methods being used to
solve the Boltzmann equation in different relic density computation packages.

Our working assumption is that the WMAP data on relic density 
constrains the current model without affecting other cosmological 
parameters in a significant way.

\subsection{Direct Detection}

Here we need to calculate the elastic cross section for
\begin{equation}
\chi {\mathcal N} \longrightarrow \chi {\mathcal N} \;,
\end{equation}
where $\mathcal N$ is a nucleus with $Z$ protons and $(A-Z)$ neutrons. 
For a Higgs mass around 125 GeV, it is heavy enough to be integrated out 
to give an effective local interaction between the dark matter and the quarks.
Since the local velocity of the dark matter is about $v_\chi \sim 10^{-3} c$,
non-relativistic reduction is appropriate. 
For the present model there is no spin-dependent cross section for the above
elastic scattering, because the dark matter is a scalar particle.
The spin-independent cross section at zero recoil energy can be obtained as
\begin{equation}
\sigma^{\rm SI}_{\chi \mathcal N} (0) =
\frac{1}{4 \pi} \mu^2_{\chi \mathcal N} 
\vert Z f_p + (A-Z) f_n \vert^2 \;,
\end{equation}
where $\mu_{\chi \mathcal N} = m_\chi m_{\mathcal N} / (m_\chi + m_{\mathcal N})$
is the reduced mass for the $\chi \mathcal N$ system and
\begin{equation}
f_N = \rho \frac{m_{N}}{m_\chi m_H^2}
\left\{ 
\sum_{q=u,d,s} f^{(N)}_{Tq} \, + \, \frac{2}{27}  \, n_Q \, f^{(N)}_{TG}
\right\}
\end{equation}
with $n_Q$ denotes the number of heavy quarks and $N = p$ or $n$.
The factors $f^{(N)}_{Tq}$ and $f^{(N)}_{TG}$ are hadronic matrix elements, and 
we will use their default values given in 
MicrOMEGAs \cite{micromegas}. 

For direct detection we choose the truncated
Maxwell velocity distribution, which is the default choice in MicrOMEGAs.
Because XENON100 is a counting experiment, the best choice for the likelihood function is 
Poisson distribution, 
\begin{equation}
\label{eq:xenonlike}
 \mathcal{L}_{\rm direct} \propto \frac{e^{-(s + b)}\left(s + b \right)^{o}}{o !} \; . 
\end{equation}
The $b = 1.0$ and $o = 2.0$ are the number of background and
observation events taken from XENON100 \cite{xenon2012},
respectively.  In order to achieve the minimum of
  $\chi^{2}_{\rm{direct}}$ equals to zero, we normalized
  Eq. (\ref{eq:xenonlike}) by the factor $e^{-o} \left( o \right)^{o}
  / {o !}$. For simplicity we do not take the background
uncertainties into account in our analysis.  The signal $s$ equals
$\varepsilon\times N(\rm{unbiased})$ 
in the nuclear recoil energy range of 6.6 - 30.5 ke$\rm{V}_{nr}$, 
where $\varepsilon$ is the
detector efficiency once the experimental cuts are applied to the
total number of unbiased events $N({\rm unbiased})$. In other words,
$\epsilon$ is the fraction of $N(\rm{unbiased})$ generated by the
Monte-Carlo simulation which survives the various cuts taken in
XENON100.

\begin{figure}[t!]
\centering
\includegraphics[width=4in]{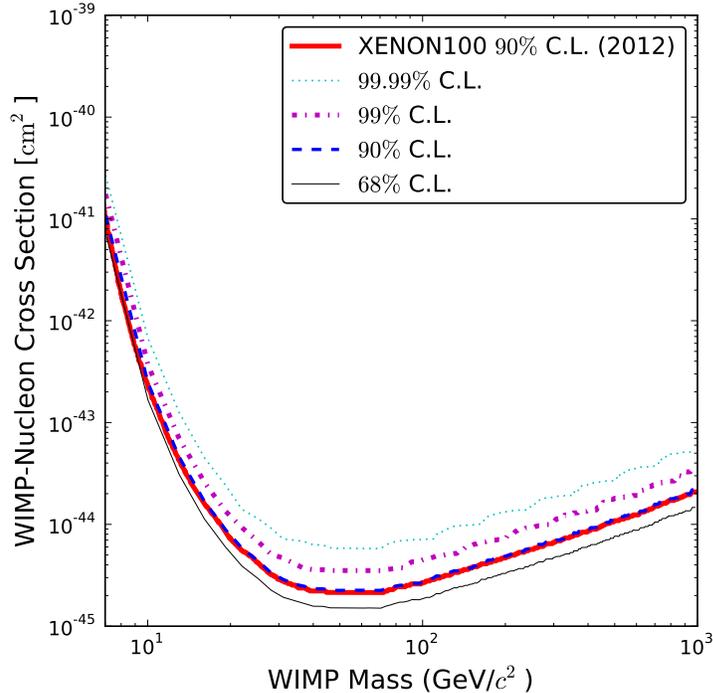}
\caption{\label{XENONlikeMAP} 
Our approximation of the XENON100 likelihood map as described in the text. 
The dashed blue line shows the 90.0\% CL bound. It approximately equals to the XENON100 
90\% CL exclusion contour, shown by the thick solid red line. 
The thin solid black line, the dashed magenta line, and the cyan dotted line 
show our calculations of the 68.0\% CL, the 99.0\% CL, and 99.99\% exclusion bound, respectively. 
}
\end{figure}

We use the following approximate method to evaluate the efficiency
$\varepsilon$ as a function of $m_\chi$.  First, from XENON100 data of
$b=1.0$ and $o=2.0$ at 90\%C.L. one can obtain a 
reference $s^*=5.72$ using the likelihood function just mentioned.  Then,
taking the values of $m_{\chi}$ and $\sigma^{\rm{SI}}_{\chi p}$ along
the 90\% C.L.  line from Fig. 3 in \cite{xenon2012} as input
to MicrOMEGAs, we can obtain $N(\rm{unbiased})$ which is
a function of $m_{\chi}$ and $\sigma^{\rm{SI}}_{\chi p}$.  Hence,
assuming the efficiency $\varepsilon$ depends only on $m_{\chi}$ and
$\sigma^{\rm{SI}}_{\chi p}$, it is simply given by the ratio
$s^*/N(\rm{unbiased})$.  Below when we compute 
the number of events for our model, we will multiply it by 
this $\varepsilon$ to obtain the corresponding number of signals.
For Poisson distribution we evaluate the effective $\chi^2$ according to the following expression
\begin{equation}
\label{effchisq}
\chi^2=-2\ln\mathcal{L} \; .
\end{equation}

In order to justify our simple treatment of the likelihood function
shown in Eq.~(\ref{eq:xenonlike}), we plot 
the 68\%, 90\%, 99\%, and 99.99\% C.L. curves, and compare to
the published XENON100  90\% C.L. curve in Fig. (\ref{XENONlikeMAP}).
It is clear that our 90\% C.L. curve is almost identical to the one from XENON100. 
Therefore, we can safely use the likelihood function in Eq.~(\ref{eq:xenonlike}). 

\subsection{Indirect Detection}

Since the couplings of the scalar dark matter to the SM fermions must go 
through the
Higgs exchange in the SZ model, the positron flux and the anti-proton flux are 
probably too small for indirect detection 
\footnote{However, it was shown in 
\cite{mambrini} that antimatter signals for the present model might be promising 
at the AMS experiment on the space station.}. 
We will focus on gamma-ray flux in
indirect detection \footnote{An earlier analysis of the gamma-ray signals for the present 
model can be found in \cite{yaguna}.}.
For continuum gamma-ray, the relevant processes in the model
are $f\bar f$, $W^+W^-$, $ZZ$, and $HH$ channels. The $W^+W^-$,
$ZZ$, and $HH$ will subsequently decay into quarks and charged leptons which can produce gamma-rays.
However, the $HH$ final state will be discarded 
in our later numerical analysis since micrOMEGAs does not provide the 
Pythia decay table of $\chi\chi\rightarrow HH\rightarrow \cdots \rightarrow$ gamma-rays. 
Furthermore, Fermi-LAT \cite{fermilat-dSphs}
has neither the limit for $HH$ nor $ZZ$ final state. 
Nevertheless, the $HH$ contribution to gamma rays 
is expected to be small, since the Higgs decays into 
SM fermion pairs are suppressed by the fermion mass. 
When $m_\chi > m_W$, 
the dominant contribution comes from $W^+ W^-$, and the
next largest contribution comes from the heaviest fermion pair that 
the $\chi\chi$ can annihilate into.
For lower DM mass, the dominant process
is the following annihilation 
\begin{equation}
\label{continuum}
\chi \chi \to H^{(*)} \to  q \bar q \to \pi^0 X \to 2 \gamma X
\end{equation}
where the neutral pion coming from quark fragmentation decays into
two photons.

Detection of one or more spectral lines would be the 
smoking gun signal for dark matter.
The following annihilation processes
$ \chi \chi \to \gamma \gamma \; , \; \gamma Z \; , \; \gamma H$
can give rise to discrete gamma lines.
In the SZ model, only the following two processes
\begin{equation}
\label{line}
\chi \chi \to H^{(*)} \to \gamma \gamma \; , \; \gamma Z
\end{equation}
are possible \footnote{By charge conjugation, it is impossible to
construct gauge invariance operators using one single photon field strength
with arbitrary numbers of Higgs fields and partial derivatives.}.
The $\gamma Z$ final state is possible if $m_\chi > m_Z/2$.
The gamma-ray lines are located approximately at energies $E_\gamma \sim m_\chi$ 
and $m_\chi (1 - m_Z^2/4m_\chi^2)$ for the two final states $\gamma \gamma$ and 
$\gamma Z$, respectively, with corrections of order
$(v_\chi/c)^2 \sim 10^{-6}$, which is minuscule.
Since these processes are one-loop induced and therefore suppressed, 
we do not include them in the 
global fitting. Instead, we will compute these cross sections after the scan 
and compare with the Fermi-LAT limits \cite{fermilat-discrete}.

The dSphs of our Milky Way are satellite systems without active star formation
or detected gas content.  They are thus fainter and expected to be
dominated by DM due to their own gravitational binding.  Although the
expected flux of gamma-rays is not as high as the Galactic Center, these
dwarf galaxies may have a better signal-to-noise ratio.  Currently,
the most stringent upper limits on the DM annihilation cross sections
in various channels are derived by Fermi-LAT Collaboration using the
new 24 months data set with the following two improvements on their
analysis \cite{fermilat-dSphs}.  First, they performed a joint
likelihood analysis to 10 satellite galaxies which can improve their
statistical power.  Second, they included the uncertainties in the
dark matter distribution in these satellites entered in the
astrophysical $J$ factor
\begin{equation}
J ( \psi ) = \int_{\rm line-of-sight , \, \Delta \Omega} dl \, 
d \Omega \rho^2 [ l(\psi) ] \;,
\end{equation}
which is the line-of-sight integral of the squared DM density, $\rho$, 
toward an observational direction, $\psi$,
integrated over a sustained solid angle, $\Delta \Omega$. 
The gamma-rays flux is then given by
\begin{equation}
\phi ( E, \psi ) = \frac{1}{8 \pi m^2_\chi} \langle \sigma v_\chi \rangle N_\gamma (E) J ( \psi ) \;,
\end{equation}
where $\langle \sigma v_\chi \rangle$ is the velocity-averaged pair 
annihilation cross section and
$N_\gamma (E)$ is the gamma-ray energy distribution per annihilation.
Based on these two improvements, robust upper limits of 95\% C.L. on
the $\sigma v_\chi$ for the $b \bar b$, $\tau^+
\tau^-$, $\mu^+ \mu^-$, and $W^+ W^-$ channels are derived in
\cite{fermilat-dSphs}.  We will use these constraints on the diffuse
gamma-ray flux in our global fitting.
Since each limit was obtained by assuming the dominance by one single channel,
we can approximately reconstruct the upper limit suitable to our case 
by applying the same method as in Sec. 4.1 of \cite{sming}. 

In our analysis for dSphs, we adopt our likelihood function as follows
%
\begin{equation}
\label{eq:ind}
 \mathcal{L}_{\rm indirect} = {\rm{erfc}}\left(\frac{\sigma v_\chi -\sigma v_{\chi \, 95}}{\sqrt{2}\tau }\right) 
 \bigg{/} {\rm erfc} \left(\frac{-\sigma v_{\chi \, 95}}{\sqrt{2}\tau }\right) \; , 
\end{equation}
where ${\rm erfc}=1-{\rm erf}$ is the complementary error function and
the effective $\chi^2$ is the same as Eq.(\ref{effchisq}).  In
addition, because the astrophysical $J$ factor is expected to have a
3\% uncertainty and the hadronization/decay tables in either
MicrOMEGAs \cite{micromegas} or DarkSUSY \cite{darksusy} have a factor
of $2$ uncertainty, we can then include the theoretical uncertainties
as $\tau=\sqrt{0.03^{2}+2^{2}}\times\sigma v_{\chi \, 95}$, where
$\sigma v_{\chi \, 95}$ is the 95\% C.L. of the reconstructed upper limit 
for our DM pair annihilation cross section.  

\subsection{Higgs Mass and Its Invisible Width}

In order to force our scan to go to the $m_{H}\sim 125$ GeV region, as
suggested by the recent LHC data \cite{atlas,cms}, we use the Gaussian
likelihood function for the Higgs mass with a central value of $125.3$
GeV and an experimental uncertainty $\sigma \sim 0.6$ GeV.  Since the
$m_{H}$ is an input, we do not introduce any theoretical error for the
Higgs mass.  Therefore, the likelihood function for the Higgs mass is
\begin{equation}
\label{eq:higgs}
{\mathcal L}_{\rm Higgs} = e^{- \frac{1}{2}\frac{\left (m_H - 125.3 \, {\rm GeV} \right)^2}{\sigma^2}} \; .
\end{equation}
It has been pointed out recently in \cite{Djouadi-etal} that 
the monojet search at the LHC has strongly disfavored 
$B_{\rm inv} \equiv \Gamma_{\rm inv} / \left( \Gamma_{\rm SM} 
+ \Gamma_{\rm inv} \right) > 0.4$ for Higgs-portal dark matter model where
$\Gamma_{\rm SM}$ is the total SM Higgs width.
The invisible width of SM Higgs in the SZ model is
\begin{equation}
\Gamma_{\rm inv} \left( H \to \chi \chi \right) = \frac{1}{32\pi}\frac{\rho^2 v^2}{m_{\rm H}}
\left( 1 - \frac{4 m_\chi^2}{m_{\rm H}^2} \right)^{\frac{1}{2}} \; .
\end{equation}
We note that the invisible decay mode is not dominant in most of dark matter mass range. 
Only when $m_{H} <130$ GeV and $m_{H} > 2 m_{\chi}$, 
the invisible decay of Higgs becomes significant.
Hence, we implement the Higgs invisible decay as 
a 0/1 hard cut. If $m_{H} <130$ GeV, $m_{H} > 2 m_{\chi}$, and $B_{\rm inv}(H\rightarrow \chi\chi)>0.4$, 
we multiply $\mathcal{L}_{\rm Higgs}$ by 0, otherwise by 1.

\subsection{Parameter Scan}

\begin{figure}[t!]
\centering
\includegraphics[width=4in]{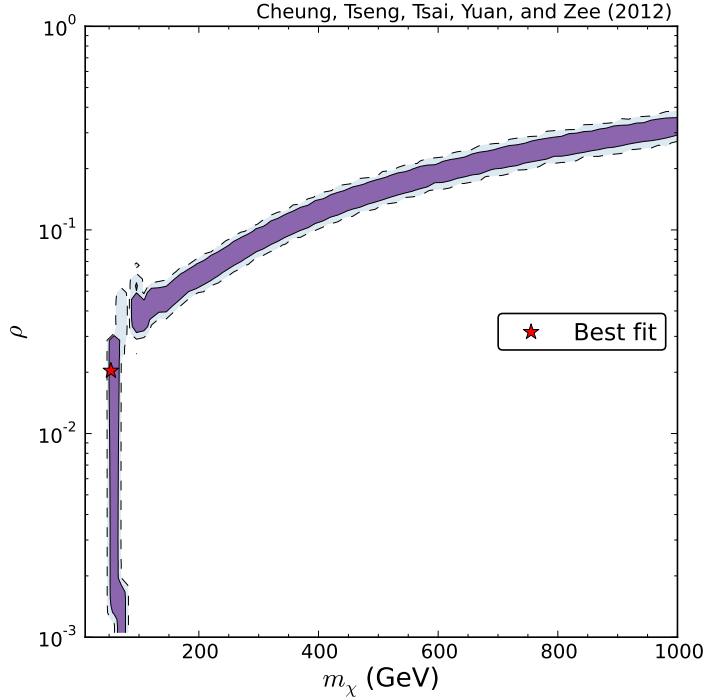}
\caption{\label{SP_mx_rho_like} 
The profile likelihood of $(m_\chi, \rho)$ for the SZ model by the global fitting using
WMAP relic density \cite{wmap}, XENON100 \cite{xenon2012}, dSphs \cite{fermilat-dSphs} 
and a 125 GeV Higgs \cite{atlas,cms} with an invisible branching 
ratio less than 40\%.
}
\end{figure}

Engaging with MultiNest v2.7 \cite{bayesian} with 10000 living points, a stop tolerance factor 0.001, 
and an enlargement factor reduction parameter 0.5, 
we perform a random scan in the three dimensional parameter space of 
$m_{\chi}$, $\rho$, and $m_{H}$ restricted in the following ranges
\begin{eqnarray}
\label{domain}
1.0 &\leq \log_{10}[m_{\chi}/\textrm{ GeV}] \leq& 3.0\nonumber\\
-3.0& \leq \;\;\; \log_{10}[\rho] \;\;\; \leq & 0.0 \\
114.0\textrm{ GeV}&\leq \;\;\; \; m_{H} \;\;\; \;  \leq & 130.0\textrm{ GeV} \nonumber
\end{eqnarray}
The selected scan range of $\rho$ is much smaller than 
the theoretical limit $|\rho|<8\pi$ because the WMAP window
is very small which only allows $\rho<1$. 
Furthermore, in order to scan efficiently in the Higgs resonance 
region and cover the low $\rho$ region, we use the log priors for 
$m_{\chi}$ and $\rho$ as specified in Eq.(\ref{domain}).
Similar results are found for the case of negative $\rho$ and will not be shown here.

After hitting the stop criteria, we collect total 440682 samples, 
and plot 68\% and 95\% profile likelihood confidence limit 
contours based on 138017 samples which are selected by Nested Sampling algorithm \cite{sampling}.
The 68\% and 95\% 
confidence limit means that the total likelihood is 
greater than $0.32*\mathcal{L}(\rm{Best\,Fit})$ and 
$0.05*\mathcal{L}(\rm{Best\,Fit})$, respectively.

\begin{figure}[t!]
\centering
\includegraphics[width=4in]{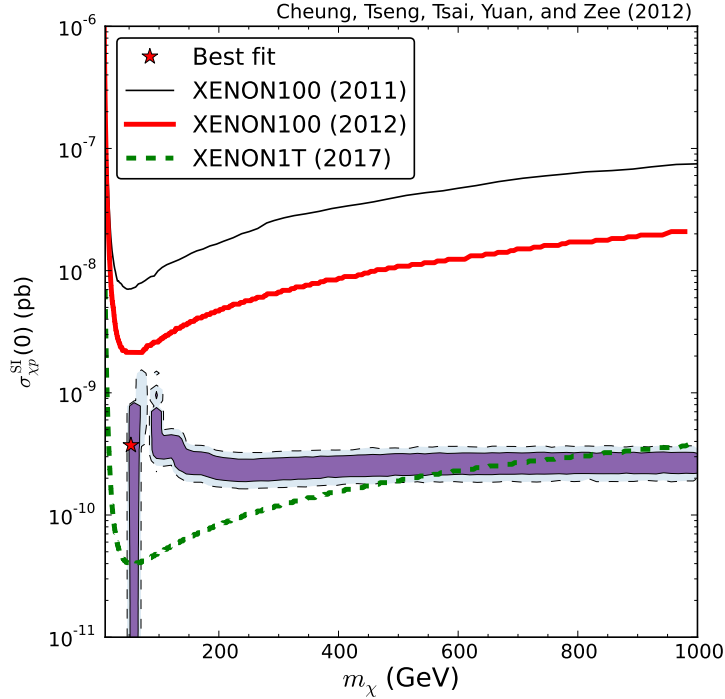}
\caption{\label{SP_mx_sigsip_like}
The profile likelihood of the spin-independent cross section 
$\sigma_{\chi p}^{\rm SI}(0)$ for the SZ model
projected onto the $m_\chi$ axis. 
The latest XENON100 limits \cite{xenon2012} are overlaid for comparison.
The projected XENON-1T sensitivity is also shown.
}
\end{figure}

The total likelihood function for our global fitting will be taken as
\begin{equation}
\label{eq:total}
\mathcal{L}_{\rm tot} = \mathcal{L}_{\rm relic} \times \mathcal{L}_{\rm direct} \times \mathcal{L}_{\rm indirect} \times \mathcal{L}_{\rm Higgs} \; ,
\end{equation}
and the effective total $\chi^2_{\rm tot}$ is given by 
\begin{equation}
\chi^2_{\rm tot} = -2 \ln \mathcal{L}_{\rm tot} \; .
\end{equation}
Our analysis uses the method of maximum likelihood. The likelihood
function of each experiment is listed clearly in Eq. (~\ref{eq:rd})
for relic density, in Eq.~(\ref{eq:xenonlike}) for the 
XENON100 data, in Eq.~(\ref{eq:ind}) for the gamma-ray data of Fermi-LAT,
and in Eq.(\ref{eq:higgs})  that for the Higgs boson mass. The joint
likelihood is then the product of all these likelihood functions, 
as given in Eq.~(\ref{eq:total}).
The ``best fit" point in Fig.~\ref{SP_mx_rho_like} presented below (as well as  in  
Fig.~\ref{SP_mx_sigsip_like}) is the point in the parameter space such that
the joint likelihood function is maximum there. The $1\sigma$ and $2\sigma$
regions in these figures are the 1$\sigma$ and 2$\sigma$ deviations 
relative from the ``best fit" point.

The result of the profile likelihood projected on the $(m_\chi, \rho)$ plane 
is shown in Fig. (\ref{SP_mx_rho_like}). 
We can clearly see that there are two branches:
the vertical branch at low $m_\chi$ region and the horizontal branch hooked around at $m_\chi >$100 GeV.
The shape of these two branches is
 mainly due to the relic density constraint.
However, XENON100 and dSphs also play a significant role
at the junction of the two branches, 
$\rho \approx 0.04 - 0.1$ and $50<m_\chi/\rm{GeV}<200$,
where relatively large $\sigma^{\rm SI}_{\chi p}$ and $\sigma v$
can be easily produced.
Furthermore, the hard cut due to the Higgs invisible 
branching ratio can remove some of the parameter space points
with $50<m_\chi/\rm{GeV}<100$ and $ 0.03< \rho< 0.1$. 
On the other hand, it is hard to satisfy our constraints
in the region $m_\chi<50 \, \rm{GeV}$, 
because the $\chi^2$ in this region rises sharply 
due to the Higgs boson mass and relic density constraints.
The vertical branch in the figure is mainly due to 
the Higgs resonance effect, 
which can efficiently enhance the dark matter annihilation cross section
when $2 m_\chi$ falls near $m_H$. Hence, the coupling $\rho$ has to be small 
correspondingly, in order to be consistent with WMAP data. 
On the other hand,
when $m_\chi >m_W$, the $\chi  \chi \to W^+ W^-$ channel 
dominates the annihilation cross section \cite{mcdonald,burgess,He:2010nt}.
Therefore, we can see from 
the figure that in 
the 1 and 2 $\sigma$ C.L. bands of the horizontal branch
the allowed $\rho$ is roughly proportional to $m_{\chi}^{2}$ (see 
Eq.(\ref{sigmavVV}) at the Appendix).

In Fig.~\ref{SP_mx_sigsip_like}, we show the profile likelihood on
$m_\chi$ - $\sigma^{\rm SI}_{\chi p}(0)$ panel against the experimental
90\% C.L. upper limit from XENON100.  Clearly, the XENON100 data is
only able to rule out 
$50 \, {\rm GeV}  \alt m_\chi \alt 100 \, {\rm GeV}$.
Current DM direct detection cannot constrain most of the parameters.
On the other hand, the Higgs resonance region and most of the 
horizontal band can be tested in the future by XENON-1T (see the dashed
line in Fig.~\ref{SP_mx_sigsip_like}).

Other than the Higgs resonance region,
the WMAP constraint dominates the likelihood function as
shown in Fig. (\ref{SP_mx_sigsip_like}), and therefore 
the largest likelihood of XENON100 only occurs at $s \ll b$. 
Nevertheless, it is easier to satisfy the relic density 
constraint in the Higgs resonance region, and therefore the
largest likelihood of XENON100 in the Higgs resonance region 
occurs at $s=1.0$ such that $s+b=o$,
by fine-tuning $m_\chi$, $\rho$, and $m_H$.
As a consequence, the best fit of our scan appears in the Higgs resonance
region.

\section{Phenomenology}

With the result of the likelihood determined, we can proceed
to evaluate other observables as predictions for the model, 
including gamma-ray lines, collider signatures, and muon
anomalous magnetic dipole moment.

\subsection{Gamma-Ray Lines} 

\begin{figure}[t!]
\centering
\includegraphics[width=3in]{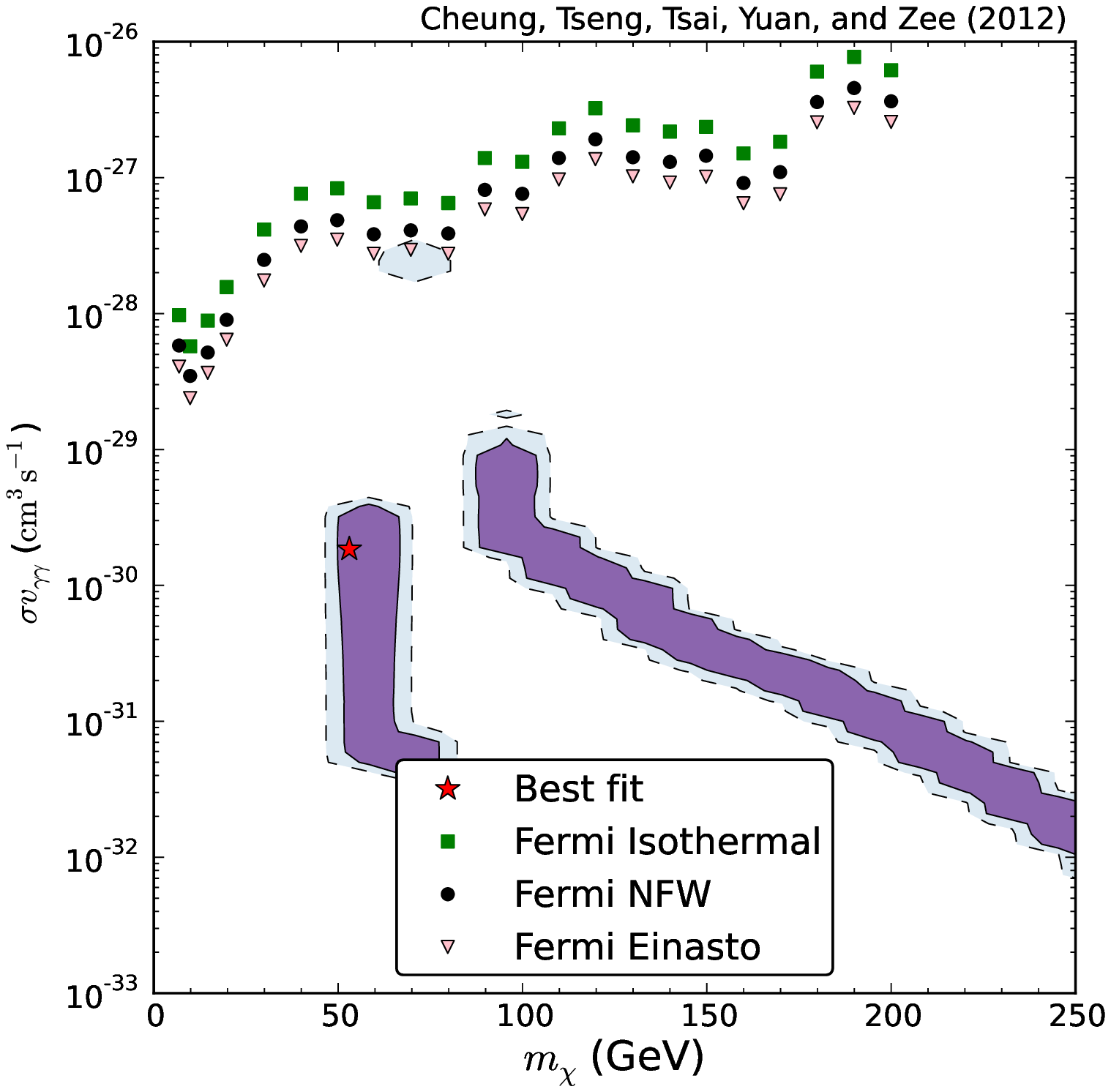}
\includegraphics[width=3in]{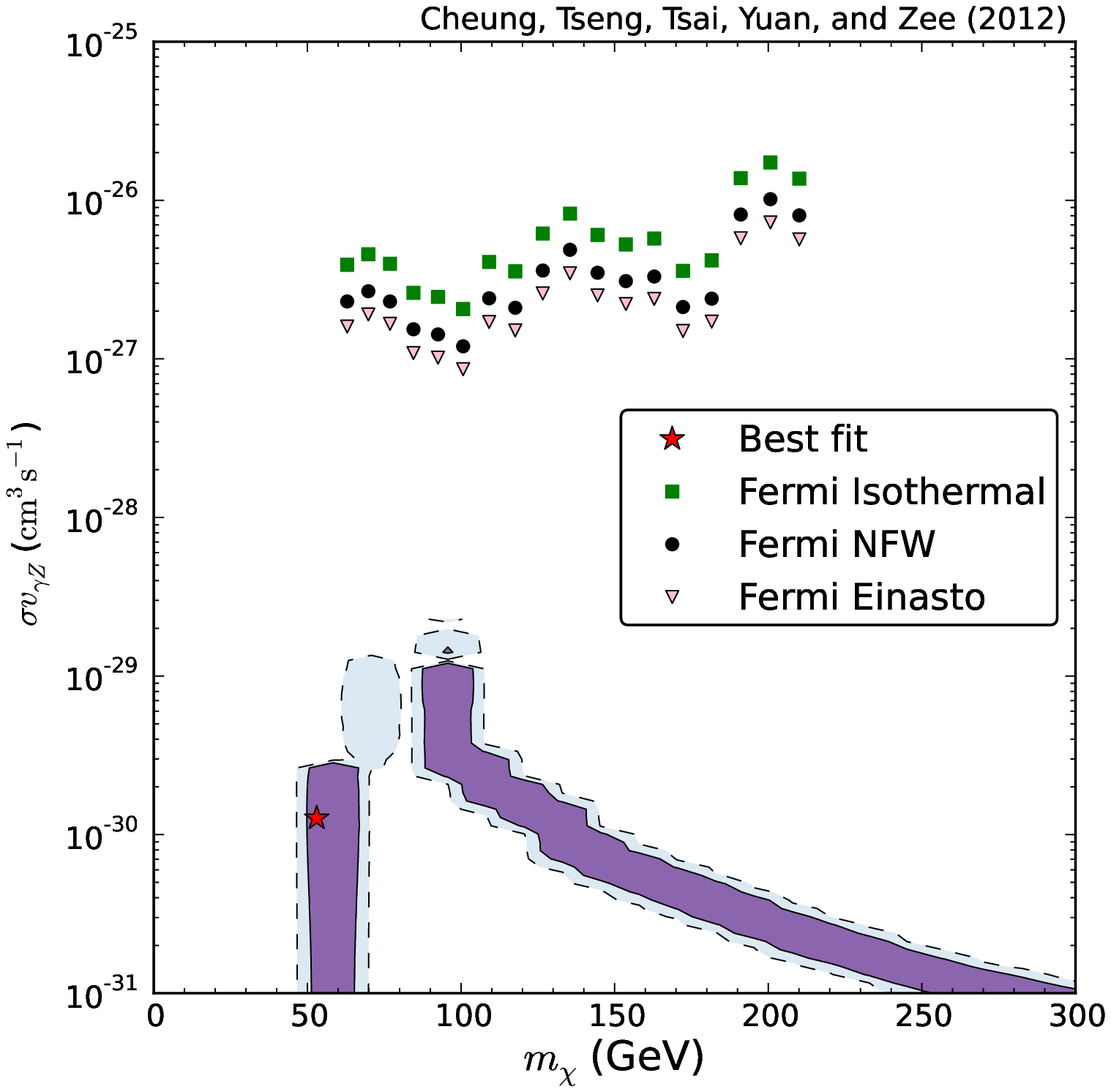}
\caption{\label{SP_mx_sigAAZ}
The annihilation cross sections for the gamma-ray line 
from $\chi \chi \to \gamma \gamma$ (left) and
$\chi \chi \to \gamma Z$ (right).
}
\end{figure}

In Fig.(\ref{SP_mx_sigAAZ}),
we plot the cross sections for the gamma-ray line in the SZ model 
versus the profile likelihood projected onto the $m_\chi$ axis.
The left panel is for 
$\chi \chi \to \gamma \gamma$ while the right one is 
for $\chi \chi \to \gamma Z$. 
The Fermi-LAT data \cite{fermilat-discrete} associated 
with different halo profiles are also shown for comparisons.
It is clear to see that the prediction for the 
$\chi \chi \to \gamma \gamma$ annihilation cross section allowed 
by the profile likelihood is well below the Fermi-LAT data
while that of $\chi \chi \to \gamma Z$ is even further below 
the Fermi-LAT data. Hopefully, future better measurements
made by Fermi-LAT can put a dent in the allowed profile likelihood.

\subsection{Collider Signatures}

If the invisible mode of $H \to \chi\chi$ opens up, 
we should study its impact on Higgs 
search at the LHC; in particular its effect on the branching ratios of 
$H \to \gamma\gamma$, $H \to W W^*$ and $ZZ^*$, which apparently show
some excesses over the background. Since the current CMS and ATLAS
data \cite{atlas,cms} showed that the excesses seen in $\gamma\gamma$,
$WW^*$, and $ZZ^*$ channels are consistent with the expectation
of the SM Higgs boson of 125 GeV,
\footnote
{The $WW^*$ and $ZZ^*$ decay modes
are slightly below while the $\gamma\gamma$ mode is somewhat higher 
than the SM predictions.} 
we cannot allow the invisible decay mode to be too large; 
otherwise the visible mode would become inconsistent
with the current data.  

It is easy to show that the branching ratio for a
visible mode would be its SM branching ratio multiplied by $(1- B_{\rm inv})$
where $B_{\rm inv}$ is the invisible branching ratio defined earlier as 
$\Gamma_{\rm inv}/(\Gamma_{\rm SM} + \Gamma_{\rm inv} )$.
In our scan in the previous section, we had required the invisible 
branching ratio 
$B_{\rm inv} < 0.4$ such that each visible mode is reduced by an amount
less than 40\% so as not to upset the current data. If the dark matter mass 
$m_\chi > m_H/2$, the Higgs boson simply
behaves like the SM Higgs boson. 

From our scan result in Fig.~\ref{SP_mx_rho_like}, a few typical points
which have the likelihood within 68\% C.L. (1-$\sigma$ band) 
can be identified as follows:

\begin{enumerate}

\item{Point A}: 
$m_\chi = 53$ GeV, $\rho = 0.02$, $m_H = 125.3$ GeV. 
The invisible branching ratio is right at $0.4$.  The significance of this
point is that the Higgs boson still has a large branching ratio into 
$\chi\chi$.  The collider signature that we will discuss below consists
of a large missing energy. 
This is the point with the maximum likelihood, shown by the star in 
Fig.~\ref{SP_mx_rho_like}.

\item Point B: 
$m_\chi = 84.0$ GeV, $\rho = 0.042$, $m_H = 125.2$ GeV.
This point gives $\chi$ a mass close to $m_W$ and hence above the Higgs decay 
threshold. This is the point at the low end of the second branch.

\item Point C: $m_\chi = 608.3$ GeV, $\rho = 0.189$, $m_H = 125.3$ GeV. This 
point gives a heavy $\chi$ that is still consistent with direct detection
limits. 

\end{enumerate}

The most common search modes so far for the dark matter are the
monojet and monophoton plus missing energies. In this model,
monojet or monophoton production must go through the Higgs boson $H$, so
that the only sizable production cross sections have to go via
\[
 b \bar b \to H \to \chi \chi\;, \qquad gg \to \chi \chi \;,
\]
in which we can attach a gluon to the $b$ or $g$ leg, or attach
a photon line to the $b$ leg. Since the $b$-parton luminosity is small
and gluon-fusion is a loop process, the monojet or monophoton rate
would be relatively small.

\begin{figure}[t!]
\centering
\includegraphics[width=3in]{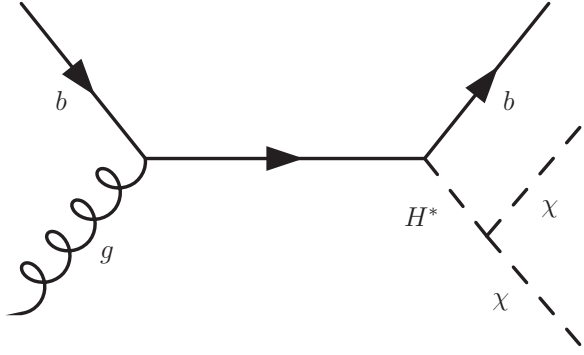}
\caption{\label{fey} A Feynman diagram showing mono-$b$ jet production
with missing energies.}
\end{figure}

Since the DM candidate $\chi$ only couples to the SM particles via the
Higgs boson, the $\chi$ will preferably couple to the heaviest fermion.
At hadronic colliders, one of the interesting processes is 
\begin{equation}
\label{b}
 g b \to b H^{(*)} \to b \chi \chi \;\; ,
\end{equation}  
where the superscript $(*)$ on the Higgs boson denotes that the Higgs 
boson could be on- or off-shell depending on the mass of $\chi$. 
Obviously, it is dominated by on-shell Higgs boson 
for $m_H > 2 m_\chi$.
If $m_H > 2 m_\chi$ and $m_H$ is lighter than $2m_W$, the Higgs boson
will dominantly decay into a pair of $\chi$.  The corresponding collider
signature would be a mono-$b$ jet plus missing energies. 
The other possible signatures would be the associated production of the 
Higgs boson with a gauge boson $W$ or $Z$:
\begin{equation}
\label{l}
 pp \to W\;(Z) H^{(*)} \to \ell \nu \; (\ell^+ \ell^-) \chi \chi \; .
\end{equation}
The final state in this case would consist of a charged lepton or a pair of
charged leptons plus missing energies.

We calculate the event rates of mono-$b$ jet, single charged lepton,
and a pair of charged leptons plus missing energies at the LHC-7, LHC-8, 
and LHC-14. We impose the following selection cuts for the $b$ jet or
charged leptons and the transverse missing energy
\begin{equation}
\label{cuts}
p_{T_b} > 30\;{\rm GeV} \, , \;\;\;\; |\eta_b | < 2 \; ; \;\;\;
p_{T_\ell} > 25 \;{\rm GeV}\,,\;\;\; |y_\ell | < 2 \; ; \;\;\; 
\not\!{p}_T > 50 \;{\rm GeV} \; .
\end{equation}
The cross sections for the mono-$b$ jet, single or a pair of charged lepton
plus missing energies are tabulated in Table~\ref{x-sec}. The largest 
cross section comes from mono-$b$ jet production. However, when we apply
the $\not\!p_T > 50$ GeV cut the cross section
mono-$b$ goes down 50 times. 
After further imposing the $B$-tagging, the event rate would only
be handful. 
Another interesting signature is the single charged lepton plus missing
energies. Counting both negatively- and positively-charged leptons the
cross section could be as high as 16 fb at the LHC-8. 
Given the LHC-8 can accumulate 20 fb$^{-1}$ each experiment, it would be
more than 300 events each experiment. The $ZH$ production would give,
on the other hand, two charged lepton plus missing energies with a few times
smaller event rates.

\begin{table}[th!]
\caption{\small \label{x-sec}
Cross sections for mono-$b$ jet, single charged lepton or a pair of
charged leptons plus missing energies arise from Higgs boson production
followed by $H \to \chi \chi$. We used the point A 
($m_\chi = 53$ GeV, $\rho = 0.02$,
 and $m_H = 125.3$ GeV, $B_{\rm inv}(H \to \chi\chi)=0.4$). 
The selection cuts are defined in  Eq.~(\ref{cuts}).
}
\begin{ruledtabular}
\begin{tabular}{lccc}
           &  \multicolumn{3}{c}{Cross sections (fb) } \\
Subprocess  &  LHC-7 & LHC-8  & LHC-14 \\
\hline
$ g b \to b H \to b \chi\chi$ &  4.6  & 6.3 &  10.4 \\
$ u \bar d \to W^+ H \to \ell^+  \nu \chi\chi $ & 9.2 & 10 & 19 \\
$ d \bar u \to W^- H \to \ell^- \bar \nu \chi\chi $ & 4.7 & 5.8 & 12 \\
$ q \bar q \to ZH \to \ell^+ \ell^- \chi\chi$ & 2.2  & 2.6 & 4.9
\end{tabular}
\end{ruledtabular}
\end{table}

Note that for other typical points of the model, e.g., points B and C,
the invisible decay mode of the Higgs boson is closed, and therefore
the decay is similar to the SM Higgs boson.  The process in Eq.~(\ref{b})
will then give rise to $3b$ or $b W W^*$ final states, depending
on the Higgs boson mass. The processes in Eq.~(\ref{l}) will give
one or two charged leptons plus either $b\bar b$ or $WW^*$.

The SM background for the mono-$b$ jet plus missing energy would be
similar to the current monojet search in ATLAS \cite{atlas-monojet} and CMS
\cite{cms-monojet}, but now with a $B$-tag on the monojet. The largest 
background \cite{atlas-monojet,cms-monojet} comes from $Z+j \to \nu \bar \nu +j$
and $W+j \to \ell \nu +j$ with minor contributions from $t\bar t$, single
top production, and QCD multijets when leptons or extra jets get missing
down the beam. 
On the other hand, background events with single or double charged
leptons plus large missing energy comes from $WZ \to \ell \nu \nu\bar \nu$
or $ZZ \to \ell\ell \nu\bar \nu$ with minor contributions from
$t\bar t$ and single top production. Precise estimations of these
backgrounds are beyond the scope of the present paper.

\subsection{Muon Anomalous Magnetic Dipole Moment}

\begin{figure}[t!]
\centering
\includegraphics[width=2in]{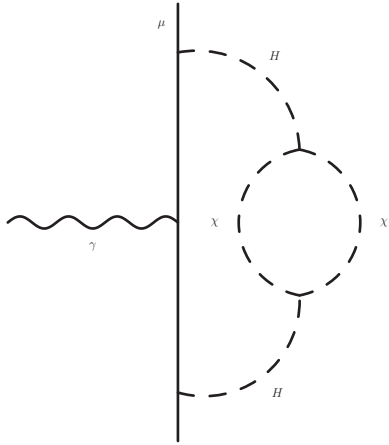}
\caption{\label{gminus2} 
Two loop Feynman diagram contributing to the muon anomalous magnetic dipole 
moment.}
\end{figure}

The experimental
value of the muon anomalous moment $a_\mu \equiv (g_\mu - 2)/2$ is
\begin{equation}
\label{gminus2exp}
a^{\rm exp}_\mu = 116 \, 592 \, 089(63) \times 10^{-11} \; ,
\end{equation}
while the SM prediction is
\begin{equation}
\label{gminus2SM}
a^{\rm SM}_\mu = 116 \, 591 \, 802(49) \times 10^{-11} \; .
\end{equation}
The $3.6 \sigma$ discrepancy between the above experimental measurement 
and theoretical calculations 
based on using the $e^+e^-$ annihilation cross section for the 
estimation of the hadronic correction
\cite{PDG}
\begin{equation}
\label{deltagminus2}
\Delta a_\mu \equiv a^{\rm exp}_\mu - a^{\rm SM}_\mu = 287(80) \times 10^{-11} 
\end{equation}
could be a harbinger of various new physics beyond the SM.
The contribution to the muon anomalous magnetic dipole moment $a_\mu$
in the SZ model first shows up at the two loop level (See
Fig. [\ref{gminus2}]). Detailed expressions can be found in the
Appendix.
In Table (\ref{muonanomaly}), we show the numerical results of $a_\mu$ for the three typical points A, B and C
from our scan.
\begin{table}[th!]
\caption{\small \label{muonanomaly}
Muon anomalous magnetic dipole moment for Points A, B and C of the likelihood.}
\begin{ruledtabular}
\begin{tabular}{ccc}
\multicolumn{3}{c}{Muon Anomalous Magnetic Dipole Moment ($a_\mu$)} \\
Point A & Point B  & Point C \\
\hline
$-1.47 \times 10^{-21}$ &  $-3.19 \times 10^{-21}$  & $-1.67 \times 10^{-21}$
\end{tabular}
\end{ruledtabular}
\end{table}
For all the relevant parameter space, we have checked that 
the contribution is negative and many orders of magnitude below the current experimental sensitivity.

\section{Conclusions}

The simplest dark matter model is realized by adding a real scalar
singlet to the standard model as was discussed quite some time ago
in \cite{SZ}, long before the popular dark matter candidate of
neutralino in MSSM model took the central stage.  In this work, we use
the most current experimental constraints of the relic density from
the 7 year WMAP data, latest XENON100 data, annihilation cross sections from
Fermi-LAT based on 10 dwarf spheroidal satellite galaxies of the Milky
Way, as well as the 125 GeV standard model Higgs candidate as discovered recently by
the LHC, to pin down the profile likelihood for the parameters $\rho$
and $m_\chi$ of the model.

The collected points are then used to evaluate the cross sections for
the gamma-ray lines from $\chi \chi \to \gamma \gamma$ and $\gamma Z$
and found that they are well under the current limits from Fermi-LAT
data.  A small part of the allowed parameter space around $m_\chi
\approx 70$ GeV barely touches the Fermi-LAT data with the Einasto halo
profile.
Recently, an interesting analysis in Ref.\cite{Weniger} using the
Fermi-LAT data suggests there could be a gamma-ray line around 130 GeV
that may be related to dark matter annihilation.  However, other authors \cite{profumo}
suggest that astrophysical sources like the fermi-bubbles \cite{fermibubbles} 
could also be responsible for this line signal.  The
gamma-ray lines in this simplest dark matter model cannot accommodate
this line signal based on the profile likelihood determined by the
global fitting with the experimental constraints mentioned above.

We also study the LHC signals of mono-$b$ jet, single charged lepton
or a pair of charged leptons plus missing energies of the model. The
most interesting case is the single charged lepton plus missing
energies which can arise from associated production of $WH$ followed
by $W \to l \nu$ and invisible decay of the Higgs. With a luminosity
of 20 fb$^{-1}$ for each experiment of ATLAS and CMS at LHC-8, we
expect several hundreds of such events based on the Point A.

We also evaluate the muon anomalous magnetic dipole moment of the
model and found that it is many orders of magnitude below the current
experimental limit for all relevant parameter space.

More stringent constraints are expected for this simple model of dark matter 
as more data from the LHC, direct and indirect detection experiments become
available in the near future.

\section*{Appendix}

\subsection*{1. Matrix Elements}

In this Appendix, we list the matrix elements and annihilation cross
sections for all the two body processes needed in the calculations of
the relic density and indirect detection.  Let $s$ to be the center of
mass energy given by $s = 4 m_\chi^2 / ( 1 - v_\chi^2/4 )$ where
$v_\chi = 2 \beta_\chi$ with $\beta_\chi$ being the velocity of the
dark matter. $N_C$ is the color factor, 1 for leptons and 3 for
quarks.

\noindent
(1) $\chi \chi \to f \bar f$:

\begin{equation}
\sum_{\rm spin} \vert \mathcal M \vert^2
=  2 N^f_C\rho^2 m_f^2 \frac{s}{(s-m_H^2)^2 + m_H^2 \Gamma_H^2} 
\left( 1 - \frac{4 m_f^2}{s} \right)
\end{equation}
\begin{equation}
\label{sigmavffbar}
\sigma v_\chi  = \frac{1}{8 \pi} N^f_C \rho^2 m_f^2 
\left( 1 - \frac{4 m_f^2}{s} \right)^{\frac{3}{2}} 
\left( \left(s - m_H^2\right)^2 + m_H^2 \Gamma_H^2 \right)^{-1}
\end{equation}

\noindent
(2) $\chi \chi \to V V$ ($V = W$ or $Z$):

\begin{equation}
\sum_{\rm spin} \vert \mathcal M \vert^2
=  \rho^2 \frac{s^2}{(s-m_H^2)^2 + m_H^2 \Gamma_H^2} 
\left[ 1 - 4 \left( \frac{m_V^2}{s} \right) + 12 \left( \frac{m_V^2}{s} \right)^2  \right]
\label{eq:chichiVV}
\end{equation}
\begin{equation}
\label{sigmavVV}
\sigma v_\chi = \frac{1}{1 + \delta_{VZ}} \frac{1}{16 \pi}
\left( 1 - \frac{4 m_V^2}{s} \right)^{\frac{1}{2}}
\rho^2 \frac{s}{(s-m_H^2)^2 + m_H^2 \Gamma_H^2} 
\left[ 1 - 4 \left( \frac{m_V^2}{s} \right) + 12 \left( \frac{m_V^2}{s} \right)^2  \right]
\end{equation}
Here $\delta_{VZ}$ is a Kronecker delta to account for the Bose statistics of the $ZZ$ final state.
We note that MicrOMEGAs computes process cross
sections by CalcHEP \cite{CalcHEP}. However, we found that the amplitude squared of
$\chi \chi \to W^+ W^-/ZZ$ differs between Eq.~(\ref{eq:chichiVV}) 
and the result from CalcHEP.  The factor inside the square
bracket of Eq.~(\ref{eq:chichiVV}) is 
$\left[ 1 - 4 \left(
      \frac{m_V^2}{s} \right) + 12 \left( \frac{m_V^2}{s} \right)^2
  \right]$, 
while the corresponding factor in CalcHEP reads 
$\left[
    \frac{m^{4}_{h}}{s^2} - 4 \left( \frac{m_V^2}{s} \right) + 12
    \left( \frac{m_V^2}{s} \right)^2 \right]$.  
Due to this discrepancy we rescale the cross section by the ratio 
of these two factors.
  
\noindent
(3) $\chi \chi \to H H$

\begin{equation}
\sum_{\rm spin} \vert \mathcal M \vert^2
= \rho^2 
\bigg\vert 
1 - \frac{3 m_H^2}{\left( s - m_H^2 \right) + i m_H \Gamma_H}
- \frac{\rho v^2}{t - m_\chi^2} 
-\frac{\rho v^2}{u - m_\chi^2}
\bigg\vert^2
\end{equation}
\begin{equation}
\sigma v_\chi = \frac{1}{64 \pi s} \left( 1 - \frac{4 m_H^2}{s}\right)^{\frac{1}{2}}
\int_{-1}^{1} d \cos\theta \sum_{\rm spin} \vert \mathcal M \vert^2
\end{equation}

\noindent
(4) $\chi \chi \to \gamma Z$

\begin{equation}
\sum_{\rm spin} \vert \mathcal M \vert^2
= \frac{\rho^2 v^2}{2} 
\frac{\left( s - m_Z^2 \right)^2}
{\left( s - m_H^2 \right)^2 + m_H^2 \Gamma_H^2}
\vert \mathcal A_{\gamma Z} (s) \vert^2
\end{equation}
\begin{equation}
{\mathcal A}_{\gamma Z} (s) = \frac{e g^2}{16 \pi^2 m_W} \left( - 4 \cos \theta_W I'_W + 
\sum_f \frac{-2 Q_f \left( T^{3L}_f - 2 Q_f \sin^2 \theta_W \right)}{\cos \theta_W} N^f_C I'_f \right)
\end{equation}
\begin{eqnarray}
I'_W &=& \int_0^1 dx \int_0^{1-x} dy 
\frac{(3 - \tan^2 \theta_W ) m_W^2 
+ xy \left[
(- 5 + \tan^2 \theta_W ) m_W^2 - \frac{1}{2} ( 1 - \tan^2 \theta_W ) s
\right]}
{m_W^2 - y(1-y)m_Z^2 + xy (m_Z^2 - s) - i 0^+} \nonumber \\
&=& (3 - \tan^2 \theta_W ) I \left( \tau_W, \tau_{ZW} \right) + \left[
(- 5 + \tan^2 \theta_W ) - 2 ( 1 - \tan^2 \theta_W ) \tau_W 
\right] J(\tau_W, \tau_{ZW})
\end{eqnarray}
\begin{eqnarray}
I'_f & = & \int_0^1 dx \int_0^{1-x} dy \frac{\left(4 xy - 1 \right) m_f^2}{m_f^2 - y (1 - y) m_Z^2 + xy (m_Z^2 - s) - i 0^+ }\nonumber \\
& = &4 J(\tau_f, \tau_{Zf}) - I(\tau_f, \tau_{Zf})
\end{eqnarray}
Here, $\tau_W = s/4 m_W^2$, $\tau_f = s/4 m_f^2$, $\tau_{ZW}=m_Z^2/4m_W^2$,
and $\tau_{Zf} = m_Z^2/4m_f^2$. $I$ and $J$ are given by
\begin{equation}
I(\tau_1, \tau_2) = \frac{1}{2} \left( \tau_1 - \tau_2 \right)^{-1} \left( f(\tau_1) - f(\tau_2) \right)
\end{equation}
\begin{eqnarray}
J(\tau_1, \tau_2) & = &-\frac{1}{8} \left( \tau_1 - \tau_2 \right)^{-1} 
\left[ 1 -  \left( \tau_1 - \tau_2 \right)^{-1} \left( f\left( \tau_1\right) - f\left( \tau_2 \right) \right) \right] \nonumber \\
&&+\frac{1}{4} \tau_{2} \left( \tau_1 - \tau_2 \right)^{-2}
\left( g\left(\tau_1\right) - g\left( \tau_2 \right) \right)
\end{eqnarray}
with $f$ and $g$ defined by
\begin{equation}
\label{f}
f \left( \tau \right) =
\begin{cases} \left[ \sin^{-1} \sqrt \tau \right]^2  & \;\; \mbox{for } \tau \leq 1 \\ 
-\frac{1}{4} \left[ \ln  \left(\frac{1 + \sqrt{1 - \tau^{-1}}}{1 - \sqrt{1 - \tau^{-1}}} \right)
- i \pi\right]^2 & \;\; \mbox{for } \tau > 1 
\end{cases} 
\end{equation}
\begin{equation}
g (\tau )= 
\begin{cases} -1 + \sqrt{\frac{1 - \tau}{\tau}} \tan^{-1}\left( \sqrt{\frac{\tau}{1-\tau}} \right) & \;\; \mbox{for } \tau \leq 1 \\ 
-1 +\frac{1}{2} \sqrt{1 - \tau^{-1}} \ln  \left(\frac{1 + \sqrt{1 - \tau^{-1}}}{1 - \sqrt{1 - \tau^{-1}}} \right)
+ \frac{1}{2} i \pi \sqrt{1 - \tau^{-1}} & \;\; \mbox{for } \tau > 1 
\end{cases} 
\end{equation}
\begin{equation}
\sigma v_\chi = \frac{1}{32 \pi} 
\left( 1 - \frac{m_Z^2}{s} \right)^{\frac{1}{2}} \frac{\rho^2 v^2}{s}
\frac{\left( s - m_Z^2 \right)^2}
{\left( s - m_H^2 \right)^2 + m_H^2 \Gamma_H^2}
\vert \mathcal A_{\gamma Z} (s) \vert^2
\end{equation}

\noindent
(5) $\chi \chi \to \gamma \gamma$

\begin{equation}
\sum_{\rm spin} \vert \mathcal M \vert^2
= \frac{\rho^2 v^2}{2} 
\frac{s^2}{\left( s - m_H^2 \right)^2 + m_H^2 \Gamma_H^2}
\vert \mathcal A_{\gamma\gamma} (s) \vert^2
\end{equation}
\begin{equation}
{\mathcal A}_{\gamma \gamma} (s) = \frac{g e^2}{16 \pi^2 m_W} \left( I_W + 
\sum_f Q_f^2 N^f_C I_f \right)
\end{equation}
\begin{equation}
I_W = 2 + 3 \tau_W^{-1} + 3 \tau_W^{-1} \left( 2  - \tau_W^{-1} \right)
f \left( \tau_W \right)
\end{equation}
\begin{equation}
\label{A_f}
I_f = -2 \tau_f^{-1} \left[ 1 + \left( 1 - \tau_f^{-1} \right) 
f \left( \tau_f \right) \right]
\end{equation}
with $f$ defined in Eq.(\ref{f}).
\begin{equation}
\sigma v_\chi = \frac{1}{64 \pi} \rho^2 v^2
\frac{s}{\left( s - m_H^2 \right)^2 + m_H^2 \Gamma_H^2}
\vert \mathcal A_{\gamma\gamma} (s) \vert^2
\end{equation}

\noindent
(6) $\chi \chi \to g g$

\begin{equation}
\sum_{\rm spin} \vert \mathcal M \vert^2
= \frac{\left( N_C^2 - 1 \right)}{4} \frac{\rho^2 v^2}{2} 
\frac{s^2}{\left( s - m_H^2 \right)^2 + m_H^2 \Gamma_H^2}
\vert \mathcal A_{gg} (s) \vert^2
\end{equation}
\begin{equation}
{\mathcal A}_{gg}(s) = \frac{g g_s^2}{16 \pi^2 m_W} \sum_q I_q
\end{equation}
where $I_q$ is given by Eq.(\ref{A_f}).
\begin{equation}
\sigma v_\chi = \frac{1}{64 \pi} \frac{\left( N_C^2 - 1 \right)}{4} \rho^2 v^2
\frac{s}{\left( s - m_H^2 \right)^2 + m_H^2 \Gamma_H^2}
\vert \mathcal A_{gg} (s) \vert^2
\end{equation}

\subsection*{2. Muon Anomalous Magnetic Dipole Moment}

Following a similar procedure as in the QED case \cite{weinberg-qft}, one can readily obtain the following result 
for the muon anomalous magnetic dipole moment
\begin{equation}
a_\mu  \equiv \frac{g_\mu - 2}{2}  =  - \frac{3 \rho^2}{32\pi^4}
\int_0^\infty d \xi \, \xi^3
\int_0^1 dx \,  
\frac{x^2 (1-x)^2} 
{\left[  H(x) + \xi^2 \right]^4}
\int_0^1 dz \log \left[ 1 + \chi(z) \xi^2 \right] 
\end{equation}
where
\begin{eqnarray}
H (x) & = & x^2 + (1 - x) r^2_{H \mu} \; , \\
\chi (z) & = & z (1-z) r^2_{\mu \chi} \; ,
\end{eqnarray}
with $r^2_{H \mu} = m_H^2/m_\mu^2$ and $r^2_{\mu \chi} = m^2_\mu / m^2_\chi$.
Performing the $\xi$ integral, we end up with a two-dimensional integration for
$a_\mu$
\begin{eqnarray}
a_\mu  
& = & - \frac{\rho^2}{128 \pi^4} \int_0^1 dx \int_0^1 dz \frac{N(x,z)}{D(x,z)}
\end{eqnarray}
where
\begin{equation}
N(x,z) = x^2 (1-x)^2 \left[
\frac{2 \left(1-H(x) \chi(z)\right)}{H^2(x) \chi^2(z)} 
 -  \left(1-\frac{3}{H(x) \chi(z)}\right) \log \left(H(x) \chi(z)\right) \right] 
\end{equation}
and 
\begin{equation}
D(x,z)  =  H^2(x) \left(\frac{1}{H(x) \chi(z)}-1\right)^3 \;\; .
\end{equation}

\section*{Acknowledgments}
This work was supported in parts by the National Science Council of
Taiwan under Grant Nos. 99-2112-M-007-005-MY3 and
101-2112-M-001-005-MY3 as well as the
WCU program through the KOSEF funded by the MEST (R31-2008-000-10057-0).
AZ was supported by the NSF under Grant No. PHY07-57035; 
he is also grateful to the Institute of Physics of the Academia Sinica, Taiwan, for its warm hospitality.
TCY is grateful to NCTS and KITPC for their warm hospitalities.
YST was funded in part by the Welcome Programme of the Foundation for Polish Science.

\end{document}